\documentclass[%
 aip, apl,
 floatfix,
 preprintnumbers,
 reprint,
 superscriptaddress,
 amsmath, amssymb,
]{revtex4-1}

\newcommand*{\paperauthor}{I.\ J.\ \surname{Vera-Marun}}
\newcommand*{\papertitle}{Quantum Hall transport as a probe of capacitance profile at graphene edges}
\newcommand*{\affrug}{Physics of Nanodevices, Zernike Institute for Advanced Materials, University of Groningen, Nijenborgh 4, 9747 AG Groningen, The Netherlands}
\newcommand*{\affnij}{High Field Magnet Laboratory, Institute for Molecules and Materials, Radboud University Nijmegen, Toernooiveld 7, 6525 ED Nijmegen, The Netherlands}

\usepackage{graphicx}% Include figure files
\usepackage{bm}% bold math

\begin{document}

\title{\papertitle}

\author{\paperauthor}
\email[e-mail: ]{I.J.Vera.Marun@rug.nl} %\homepage[]{} \thanks{}
\author{P.\ J.\ Zomer}
\author{A.\ Veligura}
\author{M.\ H.\ D.\ Guimar\~aes}
\author{L.\ Visser}
\author{N.\ Tombros}
\affiliation{\affrug}
\author{H.\ J.\ \surname{van Elferen}}
\author{U.\ Zeitler}
\affiliation{\affnij}
\author{B.\ J.\ \surname{van Wees}}
\affiliation{\affrug}

\date{\today}

\begin{abstract}
The quantum Hall effect is a remarkable manifestation of quantized transport in a two-dimensional electron gas. Given its technological relevance, it is important to understand its development in realistic nanoscale devices. In this work we present how the appearance of different edge channels in a field-effect device is influenced by the inhomogeneous capacitance profile existing near the sample edges, a condition of particular relevance for graphene. We apply this practical idea to experiments on high quality graphene, demonstrating the potential of quantum Hall transport as a spatially resolved probe of density profiles near the edge of this two-dimensional electron gas.
\end{abstract}

%\preprint{Preprint v11}

\maketitle

Since its discovery, the quantum Hall effect \cite{klitzing_new_1980} has expanded its use from the study of two-dimensional electron gas (2DEG) physics to its application in metrology \cite{jeckelmann_quantum_2001}. Therefore, it is important to understand how it manifests in realistic nanoscale devices. Of particular interest are graphene devices \cite{janssen_graphene_2011}. Single layer graphene exhibits a half-integer version of the quantum Hall effect \cite{novoselov_two-dimensional_2005, zhang_experimental_2005}, a manifestation of the Dirac electrons in graphene following a linear dispersion relation \cite{castro_neto_electronic_2009}.

Here, we demonstrate how a simple analysis of electronic transport in the quantum Hall regime can be used as a spatially resolved probe of the inhomogeneous field-effect capacitance present near the edge of a graphene 2DEG. Previous work has visualized the spatial location of edge channels in semiconductor heterostructures by using scanning probe microscopy \cite{ahlswede_hall_2001, baumgartner_quantum_2007}, consistent with theoretical work that considered the smooth confining potential electrostatically induced in those heterostructures \cite{chklovskii_electrostatics_1992}. In contrast, for graphene devices we expect a considerable charge accumulation near the edges, because the sharp confining potential is determined by the actual graphene edge \cite{fernandez-rossier_electronic_2007, silvestrov_charge_2008}. This charge accumulation is relevant for transport in the quantum Hall regime. Nevertheless, it has not been considered up to date in experimental studies in graphene \cite{novoselov_two-dimensional_2005, zhang_experimental_2005}.

\begin{figure}[bp]
\includegraphics*[angle=0, width=85mm]{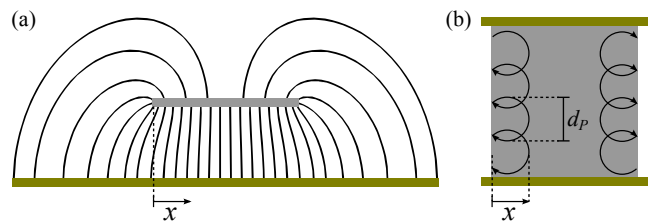}
\caption{\label{fig:scheme}
(Color online) Schematic of electric field focusing and spatial extension of edge channels. (a) Electric field lines are focused near the edges of a 2DEG (top plate), which is electrostatically doped via a gate electrode (bottom plate). (b) Semiclassical description of a two-terminal device. One possible skipping orbit with cyclotron diameter $d_P$ probes a spatial region $x$ from the edge of the device.
}
\end{figure}

To discuss the charge accumulation near the edges we consider a 2DEG electrostatically doped by applying a bias $V_G$ to a gate electrode, as shown in Fig.~\ref{fig:scheme}(a).  Electric field focusing near the edges of the 2DEG results in a position dependent field-effect capacitance per area $\alpha(x)$. Therefore, an inhomogeneous gate-induced carrier density $n(x) = \alpha(x) (V_G-V_D)$ develops along the width of the channel, with $V_D$ the charge neutrality point. We remark that the effect of the accumulated charge near the graphene edges for the simplest case of diffusive transport, without an applied magnetic field, has already been observed \cite{vasko_conductivity_2010, venugopal_effective_2011}. In contrast, we exploit the fact that in the quantum Hall regime each edge channel probes a different spatial region from the edge of the device.

With an applied perpendicular magnetic field $B$, scanning $V_G$ leads to the observation of plateaus in the Hall conductance $\sigma_H$ at quantized values of $\nu \times e^2/h$, with $\nu$ the filling factor, $e$ the elementary charge and $h$ Planck's constant. These plateaus are also observable in two-terminal devices \cite{abanin_conformal_2008}. The carrier density $n_P$ for each plateau, which depends on both $\nu$ and $B$, and the corresponding cyclotron diameter $d_P$ are given by
\begin{eqnarray}
	n_P &=& \nu \frac{e B}{h} \;, \label{eq:nu}\\
	d_P &=& 2 \sqrt{\frac{4 \pi n_P}{g_s g_v}} \frac{h}{2 \pi e B} \;, \label{eq:dc}
\end{eqnarray}
with $g_s$ ($g_v$) the spin (valley) degeneracy factor. In order to investigate the effect of the inhomogeneous capacitance $\alpha(x)$ in the quantum Hall regime, we consider a capacitance $\alpha_P$ for each individual plateau. We extract this capacitance by taking the position in $V_G$ of each plateau center together with the corresponding carrier density $n_P$, resulting in $\alpha_P = n_P / (V_G-V_D)$. The question we address is what happens with $\alpha_P$ when the field-effect capacitance $\alpha(x)$ diverges towards the edges as discussed above. To answer it we use a simple semiclassical model in which the spatial region from the edge probed by an edge channel is related to the size of the skipping orbit, ultimately limited by $d_P$, as shown in Fig.~\ref{fig:scheme}(b). Within this approximation the localization of the wave function coincides with $d_P$. We argue that a fully developed edge channel is achieved when $n(x) \geq n_P$, within a region from the edge $x = d_P$. For the inhomogeneous field-effect capacitance $\alpha(x)$ considered here, our ansatz then leads to the relation $\alpha_P \equiv \alpha(x=d_P)$. Therefore, we establish a direct mapping between the capacitance observed for each quantum Hall plateau $\alpha_P$ and the spatial location where that capacitance is probed. 

\begin{figure}[tbp] %width=85mm width=0.5\textwidth
\includegraphics*[angle=0, trim=0mm 2mm 0mm 2mm, width=85mm]{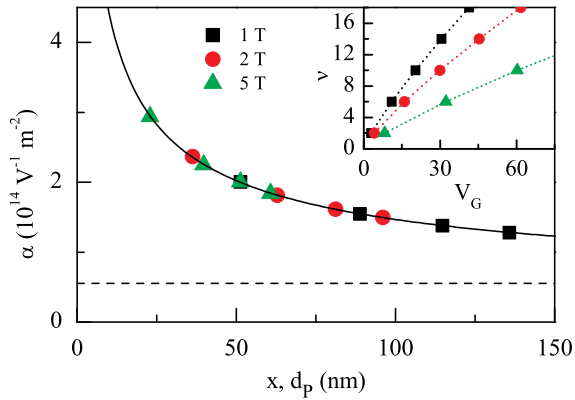}
\caption{\label{fig:theory}
(Color online) Modeling the effect of electric field focusing near the edge on the capacitance of quantum Hall plateaus, for an infinitely wide and long 2DEG suspended in vacuum $1~\mu$m over the gate. The solid line shows the calculated $\alpha(x)$ near the edge, whereas the dashed line shows the value for a simple parallel plate capacitor. Symbols correspond to $\alpha(x=d_P) \equiv \alpha_P$, for filling factors $\nu$ = 2, 6, 10, and 14 (proper for single layer graphene)\cite{castro_neto_electronic_2009} at different values of $B$. Inset: position of quantum Hall plateaus on $V_G = n_P / \alpha_P$.
}
\end{figure}

\begin{figure}[tbp]
\includegraphics*[angle=0, trim=0mm 4mm 0mm 2mm, width=85mm]{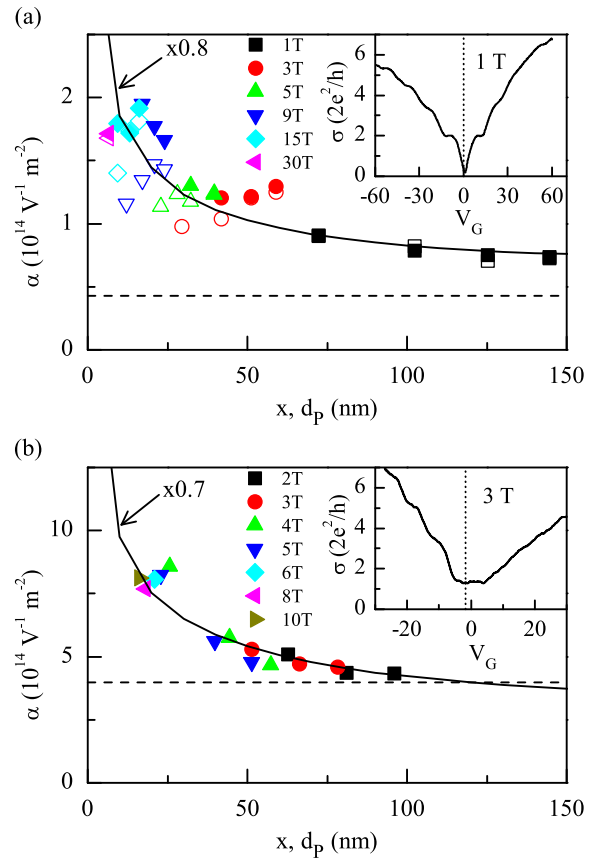}
\caption{\label{fig:data}
(Color online) Capacitance profile near the edges of high quality graphene samples.
(a) Suspended bilayer graphene, 2.6~$\mu$m long, 0.4~$\mu$m wide, with a gate dielectric of 500~nm SiO$_2$ plus 1.15~$\mu$m vacuum. (b) Single layer graphene supported on h-BN, 2.5~$\mu$m long, 2~$\mu$m wide, with a gate dielectric of 500~nm SiO$_2$ plus $\approx 40$~nm h-BN. For both (a) and (b), symbols correspond to the experimental capacitance of individual plateaus $\alpha_P$ versus $d_P$. Filled symbols are for hole transport, open symbols are for electron transport. Dashed lines show $\alpha$ for a simple parallel plate capacitor. Solid lines show the calculated $\alpha(x)$ from a three-dimensional electrostatic model averaged along the length of the graphene, scaled by a factor of 0.8 (a) or 0.7 (b). The insets show raw data for a fixed $B$ field, with the vertical dotted lines indicating $V_D$.
}
\end{figure}

Now we address some of the implications of our model for an idealized structure, a semi-infinitely wide 2DEG surrounded by vacuum and suspended at a distance of $1~\mu$m from the gate. This is a relevant configuration because it is similar to that of high quality suspended graphene devices \cite{tombros_large_2011}. We estimate $\alpha(x)$ using only electrostatics without including quantum capacitance \cite{luryi_quantum_1988}, an approach valid for our devices with thick dielectric layers \cite{fang_carrier_2007}. For this configuration one can easily treat the electrostatic problem using conformal mapping \cite{rogowski_elektrische_1923}. The calculated capacitance per area $\alpha(x)$, shown in Fig.~\ref{fig:theory}, diverges at the edge and decays towards the value for a simple parallel plate capacitor on a distance similar to the separation between the 2DEG and the gate. Note that this distance is larger than the typical cyclotron diameter in graphene devices. Following our model, we relate the capacitance profile $\alpha(x)$ to the capacitance of individual quantum Hall plateaus $\alpha_P$ represented by the symbols in Fig.~\ref{fig:theory}. For this example we consider the filling factors corresponding to the half-integer quantum Hall effect in single layer graphene. The main implication of our model is shown in the inset of Fig.~\ref{fig:theory}, where we show the position on $V_G$ of each quantum Hall plateau (for different $B$) resulting in nonlinear $\nu(V_G)$ curves. The reason why these curves are nonlinear is because the capacitance for each plateau is different, being higher for smaller $d_P$. This effect has been overlooked in previous works on high quality graphene \cite{bolotin_ultrahigh_2008, du_approaching_2008, feldman_broken-symmetry_2009} where, for a certain $B$, a constant $\alpha$ is always assumed.

In the following we use an inverse approach to that discussed above for Fig.~\ref{fig:theory}. We start from experimental quantum Hall measurements in high quality graphene devices, then extract the field-effect capacitance of individual plateaus $\alpha_P$ and finally use this data to reconstruct the capacitance profile $\alpha(x)$.

First we apply this approach to a suspended bilayer graphene sample, which is extensively studied elsewhere \cite{van_elferen_field-induced_2012}. The sample is a two-terminal device fabricated using a high yield method which allows us to obtain large mobility after current annealing \cite{tombros_large_2011}. The extracted $\alpha_P$ versus $d_P$ curve, shown in Fig.~\ref{fig:data}(a), closely follows the trend of $\alpha(x)$ expected from the focusing of electric field near the edges, as calculated by a finite-element three-dimensional electrostatic model of the sample. We observe a maximum $\alpha_P$ of $2 \times 10^{14}$~V$^{-1}$m$^{-2}$ for small $d_P$, demonstrating a threefold increase from the value at $d_P \approx 150$~nm ($0.7 \times 10^{14}$~V$^{-1}$m$^{-2}$), and a fourfold increase from the simple parallel plate capacitor model. Such an increase in capacitance confirms that our method is able to probe charge accumulation at the graphene edge. We observe similar results for another suspended sample (not shown). There is a systematic difference of $\approx20\%$ between the data and the calculation. Such a difference could be ascribed to a lower capacitance in our sample as compared to the calculated one, due to a separation between graphene and substrate which is slightly larger than the nominal value used in the calculation. This separation is given by the thickness of spin-coated polymer used to suspend graphene \cite{tombros_large_2011}.

Next, we apply our approach to a different high quality sample, a single layer graphene supported on hexagonal boron nitride (h-BN). The sample is also a two-terminal device, fabricated using a recently developed transfer technique \cite{zomer_transfer_2011}. 
For the case of hole transport, which shows higher quality with a mean free path of 110~nm, we see in Fig.~\ref{fig:data}(b) that the extracted $\alpha_P$ versus $d_P$ curve also follows the trend of $\alpha(x)$ expected from electric field focusing near the edges. We observe a maximum $\alpha_P$ of $8 \times 10^{14}$~V$^{-1}$m$^{-2}$ for small $d_P$, demonstrating a twofold increase from the value at $d_P \approx 100$~nm ($4 \times 10^{14}$~V$^{-1}$m$^{-2}$). The latter is already close to the value for a simple parallel plate capacitor model. We have observed similar results for another h-BN supported sample (not shown). There is a difference of $\approx30\%$ with the calculated $\alpha(x)$, when assuming a dielectric constant of 4 for h-BN. 
For the case of electron transport, corresponding to a low mean free path of 50~nm, we cannot resolve properly the quantum Hall plateaus except for the lowest one which does not show any appreciable increase in capacitance. 
Furthermore, we have observed that for devices using a Hall bar geometry  \cite{zomer_transfer_2011} such effect of an increased capacitance is not present, since the lateral contacts screen the electric field lines and also pin the charge density in the surrounding graphene \cite{huard_evidence_2008}.

The extracted capacitance profile for the suspended sample is considered to be a closer representation of the ideal case of using vacuum as dielectric, and possibly of its higher mobility \cite{van_elferen_field-induced_2012} (10~m$^2/$Vs at a carrier density of $2 \times 10^{11}$cm$^{-2}$, four times higher than for the h-BN supported sample) due to the current annealing process \cite{tombros_quantized_2011-1}. Scattering from defects close to the edges would lead to a larger region probed by the skipping orbits \cite{buettiker_absence_1988}, decreasing the observed charge accumulation in a similar manner as seen for the lower quality electron regime. Also, we can expect differences due to the close presence of a substrate. For example, the electrical properties of the h-BN substrate may be affected by the high electric fields present at the 2DEG edges. 
We remark that for small $d_P < 25$~nm, we observe in both type of  samples that $\alpha_P$ tends to saturate or even decreases towards the edge. This is expected because the divergence of $\alpha(x)$ towards the edge breaks down at a distance comparable to the magnetic length \cite{silvestrov_charge_2008} $l_B = \sqrt{h/2\pi e B}$. Also, at close distances from the edge the density profile depends on the exact edge structure \cite{brey_electronic_2006}. In particular, the zero-energy mode at the zigzag edge pins the Fermi energy and prevents the formation of the triangular potential well which is characteristic of sharp confinement \cite{ihnatsenka_spin_2012}. 
The aforementioned considerations are open questions which must be experimentally addressed in order to achieve a complete understanding of quantum Hall transport in realistic graphene devices.

In conclusion, we showed how to use quantum Hall edge channels to probe the physical edges in a 2DEG by extracting the capacitance profile near these edges. This practical approach allows the observation of clear differences on graphene samples with comparably high (bulk) mobilities but potentially different edge quality.

%\begin{acknowledgments}
We acknowledge B.\ Wolfs, J.\ G.\ Holstein and H.\ M.\ de Roosz for their technical assistance. This work was financed by the Zernike Institute for Advanced Materials and by the Dutch Foundation for Fundamental Research on Matter (FOM).
%\end{acknowledgments}

%\bibliographystyle{apsrev4-1}
%\bibliography{Zotero-Ivan}


\end{document}